\documentclass[reprint,superscriptaddress,amsmath,amssymb, twocolumn, floatfix,pra]{revtex4}

\usepackage{xcolor}
\usepackage{graphicx}
\usepackage{braket}
\usepackage{mathtools}
\usepackage[colorlinks, citecolor=blue]{hyperref}
\usepackage[utf8]{inputenc}
\usepackage[T1, OT1]{fontenc}

\newcommand{\mc}{\mathcal{C}}
\newcommand{\rd}{\rho_{\text{diag}}}
\newcommand{\mhc}{\mathcal{H}_c}

\DeclarePairedDelimiterX{\infdivx}[2]{(}{)}{%
	#1\;\delimsize\|\;#2%
}
\newcommand{\infdiv}{S\infdivx}

\begin{document}
\title{Characterization of anomalous diffusion in one-dimensional quantum walks}
\author{Abhaya S. Hegde}
\email{abhayas@iisc.ac.in}
\affiliation{Quantum Optics \& Quantum Information,  Department of Instrumentation and Applied Physics, Indian Institute of Science, Bengaluru 560012, India}
\author{C. M. Chandrashekar}
\email{chandracm@iisc.ac.in}
\affiliation{Quantum Optics \& Quantum Information,  Department of Instrumentation and Applied Physics, Indian Institute of Science, Bengaluru 560012, India}
\affiliation{The Institute of Mathematical Sciences, C. I. T. Campus, Taramani, Chennai 600113, India}
\affiliation{Homi Bhabha National Institute, Training School Complex, Anushakti Nagar, Mumbai 400094, India}


\begin{abstract}
Quantum walks are known to propagate quadratically faster than their classical counterparts and are used to model dynamics in various quantum systems. The spread of the quantum walk in position space shows anomalous diffusion behavior. By controlling the action of quantum coin operation on the corresponding coin degree of freedom of the walker, one can demonstrate control over the diffusion behavior.  In this work, we report different forms of coin operations on quantum walks exhibiting anomalous diffusion behavior.  Homogeneous and accelerated quantum walks display superdiffusive behavior, whereas uncorrelated static and dynamic disorders in the evolution induce strong and weak localization of the particle indicating subdiffusive and normal diffusive behavior. The role played by the interference effects in the spreading of the walker has remained elusive and our aim in this work is to present the interplay between quantum coherence and mean squared displacement of the walker. We employ two reliable measures of coherence for conclusively establishing the role of quantum interference as the driving force behind the anomalous diffusive behavior in the dynamics of quantum walks.
\end{abstract}


\maketitle
\section{Introduction} 
\label{intro}
In many complex systems, diffusion may not admit to Gaussian statistics, thereby limiting Fick's law to describe their transport behavior. A deviation from the Markovian nature of the underlying stochastic process naturally implies a corresponding withdrawal of linear time dependence on mean squared displacement, $\langle r^2(t) \rangle \sim t$. The hallmark of anomalous diffusion lies in the non-linear relationship of mean squared displacement with the time leading to a power-law pattern, $\langle r^2(t) \rangle \sim t^\alpha$, with $\alpha \ne 1$. This non-Brownian behavior is ubiquitous to a wide variety of systems ranging from liquid crystals, glasses, polymers, organelles, and even dynamics of ecosystems~\cite{HB87,BG90,BS98,GSG96,OFL19}.  Classical random walks have played an important role in modeling the dynamics in many of these complex systems. Extending such models to predict the dynamics of quantum systems necessitated the usage of quantum walks. Viewed from the lens of quantum walks, our interest here is in both cases when $\alpha > 1$, namely when enhanced diffusion prevails and also while $\alpha < 1$, underlining the extent of localization.

The origin of quantum walks bears roots in modeling the dynamics of physical particles moving on regular lattices, often termed as quantum diffusion~\cite{AAK01,Mey96}. The surge in research interest towards quantum walks, later on, was fueled by their ability to expand quadratically faster in configuration space~\cite{Amb03}. This leverage has been fruitfully exploited in the quest for improving various quantum algorithms~\cite{SKW03,Gro96,Amb03Algo,CCD03} and machine learning~\cite{MFA19}. Recently, several experiments confirm the anomalous diffusive properties of quantum walks~\cite{DKW19, GLL19}. It is only natural then to seek a characterization for their anomalous behavior. Furthermore, these walks are at the heart of quantum technology since they offer controllability in terms of experimentally tunable parameters, enabling the simulation of a variety of quantum phenomena such as transport problems~\cite{MPA08}, relativistic quantum dynamics~\cite{Str06,CBS10}, and neutrino oscillation~\cite{MMC17}.

Classical random walks are purely Markovian processes with their variance increasing linearly with the number of steps. On the other hand, their quantum analogs enjoy the features that are inaccessible to the classical realm such as quantum superposition and interference, helping in a faster (or slower) spreading. This suggests that the quantum speedup (or slowdown) is due to the arrangement of amplitudes of interfering paths and the underlying dynamics are a collective result of the constructive and destructive interference. In fact, it was analytically shown that the quadratic increase in the variance of the quantum walker is a direct consequence of quantum coherence~\cite{RSS04}. Nevertheless, a conclusive argument to establish a direct correlation between coherence and variance of a quantum walker remains to be made. Our work closely looks at exemplifying this crucial difference between the coherence of classical and quantum walks and subsequently demonstrates the role of coherence in exhibiting anomalous diffusion in quantum walks. We take the examples of homogeneous, accelerated, and disordered quantum walks in characterizing anomalous diffusion using quantum coherence.

This paper is organized as follows. We begin with an introduction to the kinds of walks considered for this study in Sec.~\ref{sec:types_of_dtqw}. The main results of our analysis on anomalous diffusion of quantum walks are outlined in Sec.~\ref{sec:anomalous_behavior}. Its relationship with coherence is extensively discussed using two suitable measures in Sec.~\ref{sec:measures}. We conclude with a broad summary of the work in Sec.~\ref{sec:summary}.

\section{Discrete-time Quantum Walks and Anomalous  diffusion}
\label{sec:types_of_dtqw}
\subsection{Homogeneous quantum walks}

Discrete-time quantum walk in one-dimensional space is  defined  on a composite Hilbert space $\mathcal{H}$ comprising of a coin degree of freedom $\mathcal{H}_c$ and a position degree of freedom $\mathcal{H}_p$ so that $\mathcal{H} = \mathcal{H}_c \otimes \mathcal{H}_p$. Here the coin Hilbert space is spanned by the two internal degrees of freedom of the walker, denoted as $\{\ket{\uparrow},\ket{\downarrow}\}$, while the position Hilbert space $\mathcal{H}_p$ has $\{\ket{x}\}$ where $x \in \mathbb{Z}$ as its basis. The dynamics of the quantum walk evolution is defined using action of quantum coin operation on $\mhc$ followed by a conditioned position-shift operation on the composite space $\mathcal{H}$.  Coin operation on walker will evolve the particle into the superposition of the internal degree of freedom. A generalized coin operator can be written as an arbitrary SU(2) matrix of the form~\cite{CSL08},
\begin{equation}
\label{coin}
\hat{C}(\xi, \eta, \theta) \equiv 
\begin{pmatrix}
e^{i\xi} \cos\theta & e^{i\eta}\sin\theta\\
-e^{-i\eta}\sin\theta & e^{-i\xi}\cos\theta
\end{pmatrix}.
\end{equation}
A rich family of walks can be unfolded by appropriate modifications to the coin operation $\hat{C}$. In this work, we consider a simpler version of the coin operator for defining the homogeneous walk, denoted by $\hat{C}(\theta) \equiv \hat{C}(\xi = 0,\, \eta = \pi/2,\, \theta)$. The coin operation is followed by the conditioned position shift operation which carries the superposition initially endowed on the internal degree of freedom to the position space of the walker. The operator conditioned on internal state of the particle will be of the form,
\begin{equation}
\hat{S}_x \equiv \sum_{x \in \mathbb{Z}} \Big(\ket{\uparrow}\bra{\uparrow} \otimes \ket{x-1}\bra{x} + \ket{\downarrow}\bra{\downarrow} \otimes \ket{x+1}\bra{x}\Big).
\end{equation}
Therefore, the unitary evolution operation for each step of the walk is given by $\hat{W}_x(\theta) \equiv \hat{S}_x \cdot \left[\hat{C}(\theta) \otimes \mathbb{I}\right]$
and after $t$-time steps, state of the system is 
\begin{equation}
\label{qw_unitary}
\ket{\psi_t} = [\hat{W}_x(\theta)]^t \ket{\psi_{0}}.
\end{equation} 
Here the initial state is,
\begin{equation}
\label{initial_state}
	\ket{\psi_{0}} = (c_1 \ket{\uparrow} + c_2 \ket{\downarrow}) \otimes \ket{x=0},
\end{equation} with $c_1^2 + c_2^2 = 1$. We choose $c_1 = c_2 = \frac{1}{\sqrt{2}}$ in the following discussions. The coin parameter $\theta$ controls the variance $\sigma^2$ of the probability distribution in the position space~\cite{CSL08} and this distribution spreads quadratically faster ($\sigma^2 \approx [1-\sin(\theta)t^2]$) in position space when compared to the classical random walk~\cite{Amb03}. 

\subsection{Disorder inducing quantum walks}
From the Eq.~\eqref{qw_unitary}, it is clear that a time-translation symmetry is inbuilt in the evolution of a homogeneous walker. Introducing a dynamic disorder by randomly changing the action of coin operator at each discrete time step breaks this symmetry and subsequently leads to a localization of the particle in position basis~\cite{BNP06, RMM04, WLK04, Cha11, YKE08, Kon10}. By the same token, one may plug in a static disorder by operating a random coin rotation at each position to severely localize the particle. While the temporal disorder in quantum walk leads to a weak localization, the spatial disorder is known to induce Anderson localization~\cite{JM10, Cha13, RAS05, VFQ17}. Such disruption from the homogeneity present in standard quantum walks could be detrimental to optimizing search algorithms. On the contrary, both of these disorders have been studied extensively in enhancing the entanglement and non-Markovianity generated between the internal and external degrees of freedom~\cite{VAR13, PBC18}. 

We consider two uniform probability distributions based on two independent identically distributed random variables, denoted by $\tilde{\theta}_t$ and $\tilde{\theta}_x$, for modeling the temporal and spatial disorders, respectively. This form of white noise borrowed from classical Markovian processes performs imperfect rotations of the form $e^{-i \tilde{\theta}\sigma_x}$. We remark here that these can be modeled as classical noises rather than those brought upon by genuine system-bath interactions.

While simulating the temporal disordered quantum walk, we have used a random $\theta_t$ at each step from the uniform distribution of $\tilde{\theta}_t$ defined over the range $[0,\, \pi]$. The final state after $t$ time-steps corresponds to $\ket{\psi_t} = $
\begin{equation}
\hat{W}_x(\tilde{\theta}_t)\ket{\psi_0} = \hat{W}_x(\theta_t)\hat{W}_x(\theta_{t-1})\cdots\hat{W}_x(\theta_{1})\ket{\psi_0}.
\end{equation}

Likewise, we define the spatial disorder in DTQW evolution by acting random $\theta_x \in [0, \, \pi]$ at each position, thereby involving a position dependent coin operation in the walk unitary~\cite{Cha13},
\begin{equation}
\hat{W}_{\tilde{\theta}_x} = \hat{S}_x \cdot \left(\sum_{x} \hat{C}(\theta_x) \otimes \ket{x}\bra{x}\right).
\end{equation}

Assuming a plane wave solution for the wavefunction $\psi_{x, t} = e^{-i(kx - \omega t)}$, the group velocity of a disordered walker would boil down to a summation of uniform random values in $[-1, 1]$ which is then averaged over either all instances of time $t$ in case of temporal disorder, or $2t+1$ positions in case of spatial disorder. While this may add up to a negligible positive number for a smaller number of steps, it is evident that in the asymptotic limit, the mean group velocity drops to zero in both cases: $\lim\limits_{t \to \infty} \langle v_g^{\text{SD/TD}} \rangle \approx 0$ leading to localization. Mean group velocity settles to zero faster for a walk with spatial disorder resulting in Anderson localization compared to temporal disorder which leads to a weak localization~\cite{Cha11, CB15}. The localization length is usually a function of the coin parameter $\theta$ given as $\xi = -\left[\ln(\cos\theta)\right]^{-1}$~\cite{VFQ17, VFF19}.

\subsection{Accelerated quantum walks}

Introducing an exponentially decaying parameter in the rotational angle of the coin results in a walk whose mean square deviation exceeds that of the homogeneous one, thereby making the dynamics `accelerated'. Denoting the coin parameter in the homogeneous quantum walk as $\theta_0$ in Eq.~\eqref{coin}, it is turned into a step-dependent operation of the form $\theta = \theta_0 e^{-at}$ with the parameter $a$ deciding how quickly the walk turns fully ballistic. Thus the spread is bounded below by $t \cos(\theta_0)$ and the wider spreads are achieved by modulating positive values for $a$. Consequently, larger $a$ values lead to a faster spread with the standard deviation attaining the maximum of $t$. As time goes by, ballistic fronts run away much faster than the spreading of the classical or even normal quantum walk diffusive evolution. This time-dependent coin operation has been previously explored in the context of enhancing the entanglement between the spin and positional degrees of freedom of the accelerated walker~\cite{SBL19, MEH18}.

\begin{figure}[!ht]
\centering
\includegraphics[width=\columnwidth]{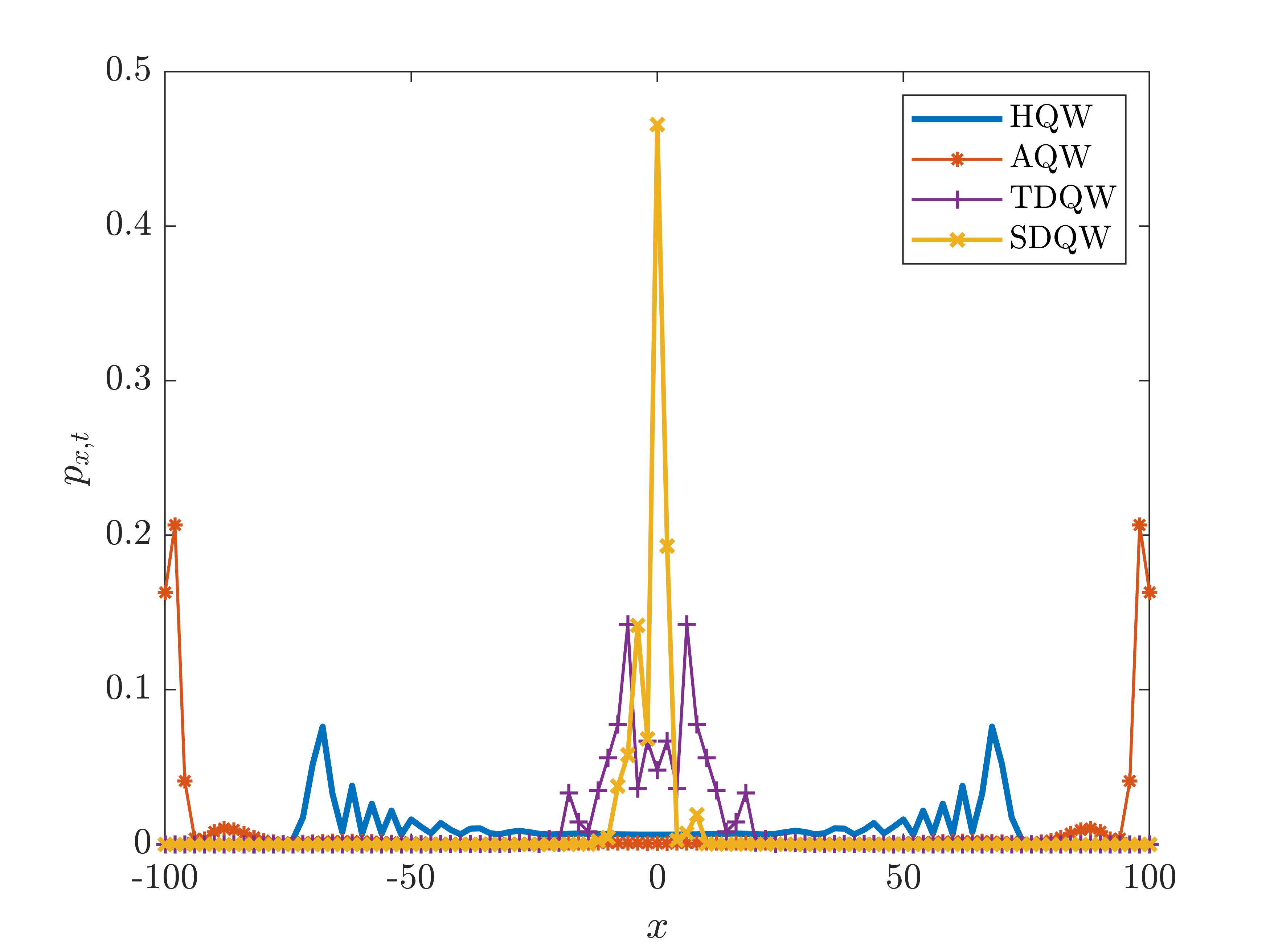}
\caption{\label{fig:probability}Probability distributions for all four walks described in Sec.~\ref{sec:types_of_dtqw} is plotted for 100 steps. Vanishing intermediate points in the probability distribution are removed. Maximum spread is visible in accelerated walk followed by the homogeneous one, while prominent localization is seen in temporal and spatial disordered walks. For homogeneous walk a coin parameter $\theta = \pi/4$ has been chosen, while keeping $\theta_0 = \pi/4$ and $a=0.02$ for the accelerated walk. To minimize the biases arising from randomness in choosing the coin parameter, an average over 100 trials was determined for temporal and spatial disordered walks. Plots in this work can be assumed to follow the same specifications for all the mentioned walks, unless specified otherwise.}
\end{figure}

The probability distributions for all these walks at a site $x$ after $t$ steps is given by
\begin{equation}
p_{x, t} = |\psi^{\uparrow}_{x, t}|^2 + |\psi^{\downarrow}_{x, t}|^2.
\end{equation}
This is plotted for all walks after 100 steps with the zeros of the distribution removed as shown in Fig.~\ref{fig:probability}. The accelerated quantum walk (AQW) with $a = 0.02$ and $\theta_0 = \pi/4$ results in a distribution with two peaks at either end of lattice, making it wider than that of homogeneous (HQW) and disordered walks. Simulated over 100 trials to weed out the biases, the spatial  (SDQW) and temporal disordered  (TDQW) walks are localized near the origin. The homogeneous discrete-time quantum walk with $\theta = \pi/4$ is neither localized nor fully ballistic. The symmetric distribution in the probability of position space is due to the equal superposition of basis states at the initial step.

\section{Anomalous behavior}
\label{sec:anomalous_behavior}

The mean square deviation (MSD) can be evaluated using the moments of position space as
\begin{equation}
	\label{msd}
	\langle r^2 \rangle = \langle x^2 \rangle - \langle x \rangle ^2, \quad \langle x^n \rangle = \sum\limits_{x} x^n p_{x, t}.	
\end{equation} 
The standard deviation $\sigma$ is then identified as the square root of MSD~\cite{MF21}.

In a one-dimensional Brownian diffusive process for an initially localized wavepacket, the mean square deviation grows linearly in time, $\langle r^2 \rangle = D t$ with $D$ being the diffusion constant. Anomalous diffusion is characterized by non-linear diffusive growth with $\alpha$ being the exponent. For the growth of $\alpha > 1$, we turn to the regime of superdiffusion as it ensures faster than classical diffusion. Conversely, an $\alpha < 1$ implies a limited movement of the walk stemming from a localization-inducing potential.

\begin{figure}[!ht]
	\centering
	\includegraphics[width=\columnwidth]{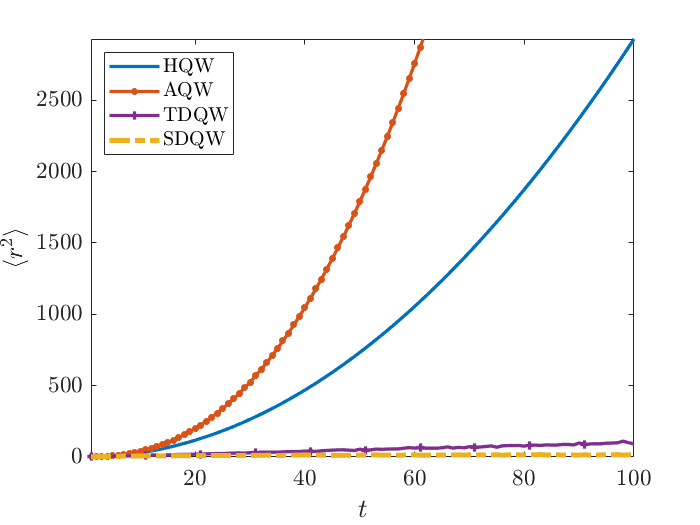}
	\caption{\label{fig:msdplot}Mean square deviation from Eq.~\eqref{msd} is plotted against the 100 time-steps for various kinds of discrete-time quantum walk explored in Sec.~\ref{sec:types_of_dtqw} with the same parameters used for plotting Fig.~\ref{fig:probability}. Homogeneous and accelerated walks are naturally superdiffusive, while the disorders impede  group velocity of the particle forcing natural diffusion and subdiffusion.}
\end{figure}

\begin{figure}[!ht]
	\centering
	\includegraphics[width=\columnwidth]{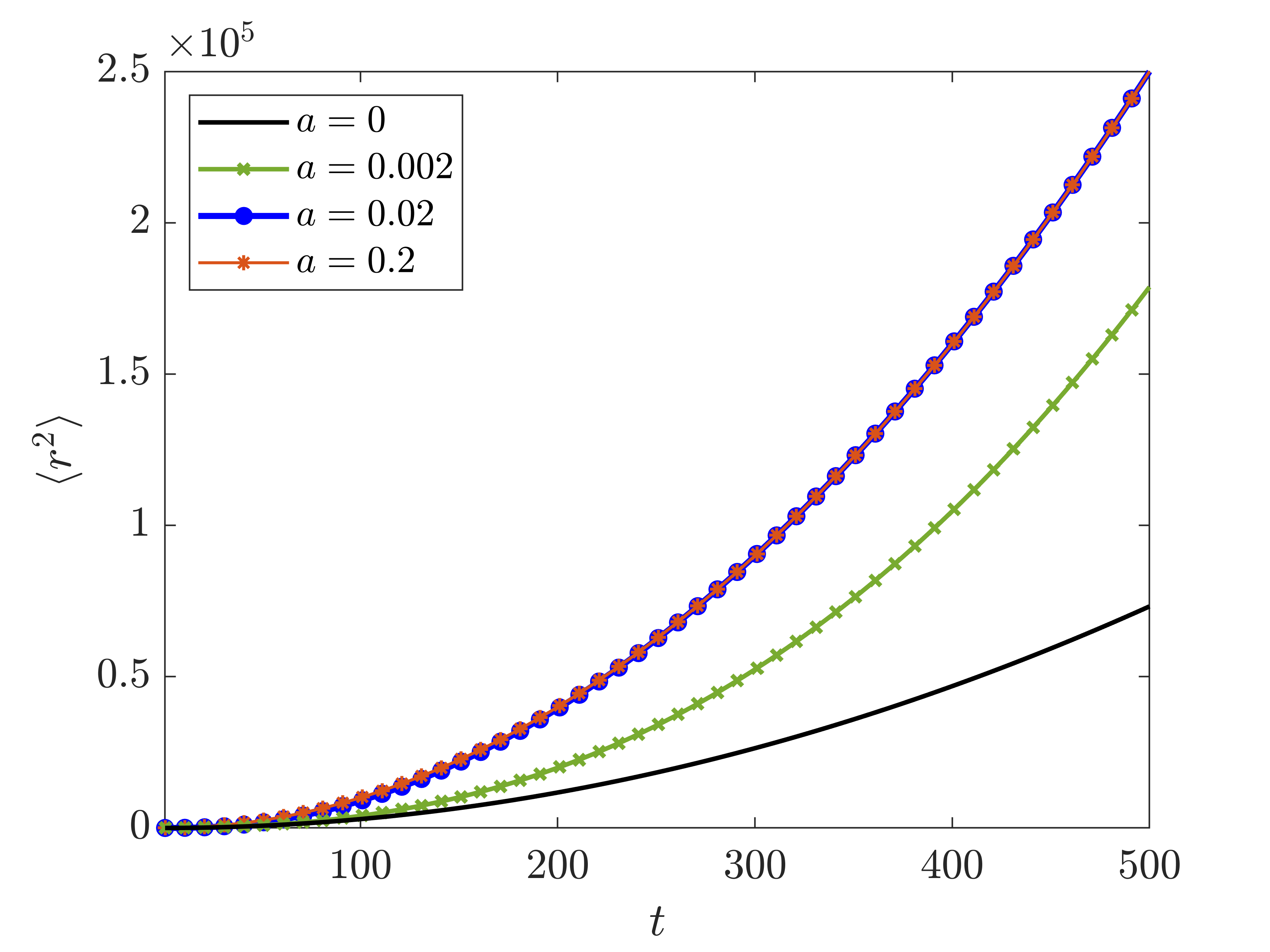}
	\caption{\label{fig:msd_acc_plot}Variance for accelerated walk is compared against homogeneous ($a=0$) for 500 number of timesteps with the parameters as specified in the legend. Modulating the values of $a$ results in varying curvature $\alpha$.}
\end{figure}

This classification for quantum walks can better be viewed in Fig.~\ref{fig:msdplot} where the growth of variance is visualized over time. After 100 steps of each type of evolution described in the previous section, the variance is seen to grow non-linearly with the number of steps. As expected from Fig.~\ref{fig:probability}, the accelerated walk leads with the highest variance, while that of the homogeneous walk comes close to it. Both of these walkers clearly maintain an $\alpha > 1$ positing them as superdiffusive. Higher values of $a$ in accelerated walk correspond to the faster interspersal of the particle in position space leading to a larger deviation from the normal quantum walk as is evident from the evolution of 500 steps in Fig.~\ref{fig:msd_acc_plot}.

The spatial disordered walks fall into the domain of subdiffusion as can be derived from Fig.~\ref{fig:msdplot}. The localization in the position space is a glaring indication of its lesser than classical variance, thereby fixing an $\alpha < 1$. The reason for the halting of spreading during evolution is due to the group velocity of waves approaching zero, as discussed previously. For finite timesteps, the variance for the walk having temporal disorder results in normal diffusion behavior identical to classical random walk but at the asymptotic limit from the point where group velocity becomes almost zero we expect a deviation towards subdiffusive behavior.

\begin{figure}[!ht]
\includegraphics[width=\columnwidth]{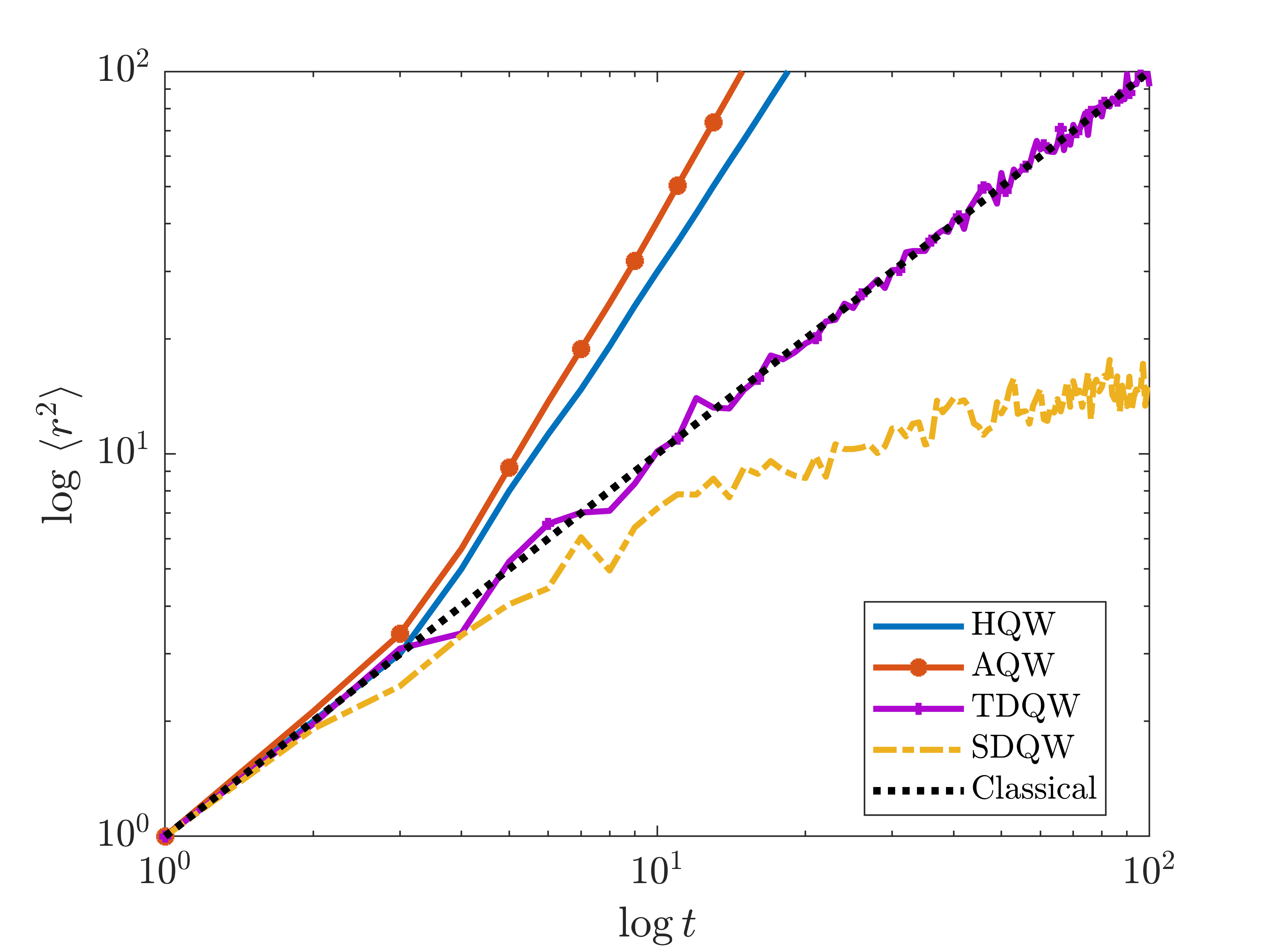}
\caption{\label{fig:alpha} Approximate $\alpha$ values for the walks in the order of legends: $1.85$, $1.64$, $0.99$, $0.68$, $1$. For the time scale we have plotted, we can seen a clear overlap of temporal disordered quantum walk with the classical random walk showing a normal diffusive behaviour.}
\end{figure}
In Fig.~\ref{fig:alpha} we have shown the logarithmic scale of MSD shown in Fig.~\ref{fig:msdplot} and the approximate values of $\alpha$ are 1.85, 1.64, 0.99 and 0.68 for accelerated, homogeneous, temporal disordered, and spatial disordered quantum walks, respectively. 

The argument for quantum walks being radically different from classical walks can be further strengthened with the following: a quantum walker whose coin or positional degree of freedom is iteratively measured after each step results in a classical random walk. Suppose we begin with a localized position state $\ket{0}$, and repeatedly perform Hadamard operation $H \equiv \hat{C}(0,\, 0,\, \pi/4)$ on the coin space, the resulting transformation after one such step will be, 
\begin{subequations}
	\begin{align}
	\ket{\downarrow} \otimes \ket{0} &\overset{H} \longrightarrow \frac{1}{\sqrt{2}} (\ket{\uparrow} - \ket{\downarrow}) \otimes \ket{0} \\
	& \overset{S} \longrightarrow \frac{1}{\sqrt{2}} (\ket{\uparrow} \otimes \ket{-1} - \ket{\downarrow} \otimes \ket{1}).
	\end{align}
\end{subequations}
A subsequent measurement on the coin space in the standard basis leaves the states $\{\ket{\uparrow} \otimes \ket{-1},\, \ket{\downarrow} \otimes \ket{1}\}$ with probability of $1/2$ each and thus destroys the interference in the position space. Continued measurements effectively reset the walker destroying the quantum advantage held in the form of correlations in position space. Thus, after $t$ such steps the distribution approaches a Gaussian centered around zero with a variance of $t$ resembling a classical random walk.  We should note that though the variance with subsequent measurement and finite time temporal disorder result in variance identical to classical random walk, in the latter case quantum interference persists.

\section{Measures of coherences}
\label{sec:measures}

One of the distinguishing features of quantum walks is the presence of quantum interference effects. The dynamics of a quantum walker in discrete-time walks is akin to a multi-path interference as it evolves in position space involving interference of amplitudes of multiple traversing paths. The absence of such effects would manifest as a purely classical walk or a directed transport behavior, as previously noted. It is natural to probe the role of interference in displaying the discussed anomalous behavior. To that end, we apply suitable quantifiers of interference and compare their behavior across the quantum walks described in Sec.~\ref{sec:types_of_dtqw}.

Central to all experimental setups involving two slits or two paths, visibility is usually deemed as the straightforward measure of coherence of waves~\cite{Gha05}. Following the need for a similar quantity to capture interference occurring in physical systems involving more number of interfering paths or particles, a number of studies found quantum coherence to be the analogous figure-of-merit~\cite{Qur19, MVQ19}. The notion of coherence is routinely harnessed in experiments of quantum optics and quantum technology~\cite{GLM11, NC10, WM94}. Here, we make use of the rigorous framework introduced for adopting quantum coherence as a physical resource~\cite{BCP14}. Based on the conditions that a valid measure of coherence must be vanishing on any incoherent set of states, monotonic under the action of incoherent quantum channels, and non-increasing under the mixing of quantum states, it was conclusively established that relative entropy of coherence $\mc_{RE}$ and $l_1$-norm based coherence $\mc_{l_1}$ were the only general measures satisfying all three criteria~\cite{BCP14}. We restrict ourselves in determining the values of these \textit{bona fide} measures of coherence for our walks of interest.

\subsection{$l_1$-norm coherence}

The $l_1$-norm based measure is a widely used quantifier of coherence given by the absolute sum of off-diagonal elements of the density matrix associated with the state of the system $\rho$: $\mc_{l_1}(\rho) = \sum_{j \ne k} |\rho_{j,k}|$ where $\rho_{j,k} = \langle j \lvert \rho \rvert k \rangle$. 
A normalized version of this definition applicable to arbitrary dimensions was later proposed as~\cite{BQS15},
\begin{equation}
\label{extendedL1}
\mc'_{l_1}(\rho) = \frac{1}{n-1}\sum_{j \ne k} \lvert\rho_{j,k}\rvert.
\end{equation}
From here onwards, we use the normalized $l_1$ coherence for further analysis with the relabeling $\mc_{l_1}' \equiv \mc_{l_1}$. The degree of quantumness is intuitively ascribed to the off-diagonal terms and the measure at hand succinctly reflects that. Furthermore, this measure meets all the criteria defined  for claiming any quantity is equivalent to visibility by~\cite{Dur01}. Accordingly, one can think of normalized coherence as an extension of visibility for multi-path interference.

The evolution of the combined system at each step of the walk is governed by the walk unitary $W$ and the resulting joint density matrix will be $\rho_t = \hat{W}^t \rho_0 (\hat{W}^{\dagger})^t$ with $\rho_{0} \equiv \rho_c(0) \otimes \rho_p(0) = \ket{\psi_{0}}\bra{\psi_{0}}$ from Eq.~\eqref{initial_state}. Since we are interested in the diffusion of wavepacket in the position space only, we trace out the coin degrees of freedom for calculating the coherence left in the positional degree of freedom, 
\begin{equation}
\rho_p(t) \equiv \text{Tr}_c[\rho_t] = \text{Tr}_c[\hat{W}^t\rho_{0}(\hat{W}^{\dagger})^t].
\end{equation}
Each step of the quantum walk unravels two sites on both ends of one-dimensional lattice, thus accumulating $2t+1$ number of positions after $t$ steps. Noting down that density matrices are Hermitian, we can simplify the Eq.~\eqref{extendedL1} to a form involving the absolute values of only upper (or lower) triangular entries,
\begin{equation}
\label{norm_coh}
\mc'_{l_1}(\rho_p) = \frac{1}{t}\sum_{j>k} \big|(\rho_p)_{j,k}\big|.
\end{equation}
Note that this equation is proportional to $1/t$ and hence coherence values will eventually precipitate to zero. Thus it is enough to overview the normalized coherence as in Eq.~\eqref{norm_coh} for the first few steps, say 100, for establishing the role of interference in spreading. 

\begin{figure}[htb]
\centering
\includegraphics[width=\columnwidth]{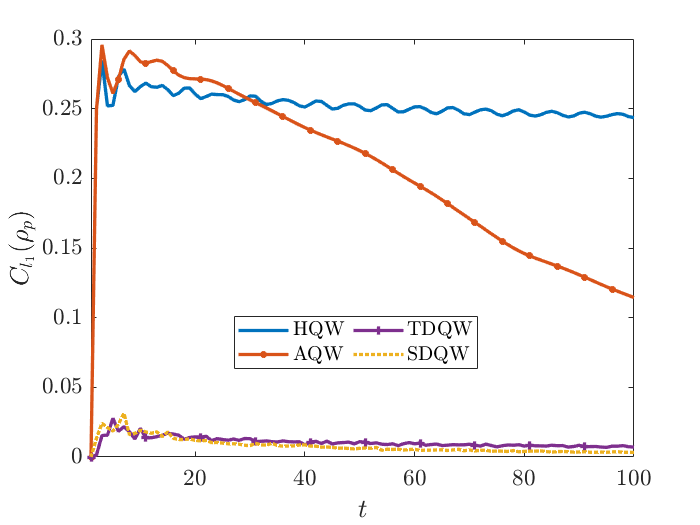}
\caption{\label{fig:coherenceL1}$l_1$-norm based coherence for position space is plotted against the number of steps for all the walks described in Sec.~\ref{sec:types_of_dtqw}. The parameters are fixed at the same value as used for Fig.~\ref{fig:probability}. Homogeneous and accelerated walks display higher coherence in comparison to disordered walks, just as expected. }
\end{figure}

\begin{figure}[htb]
	\centering
	\includegraphics[width=\columnwidth]{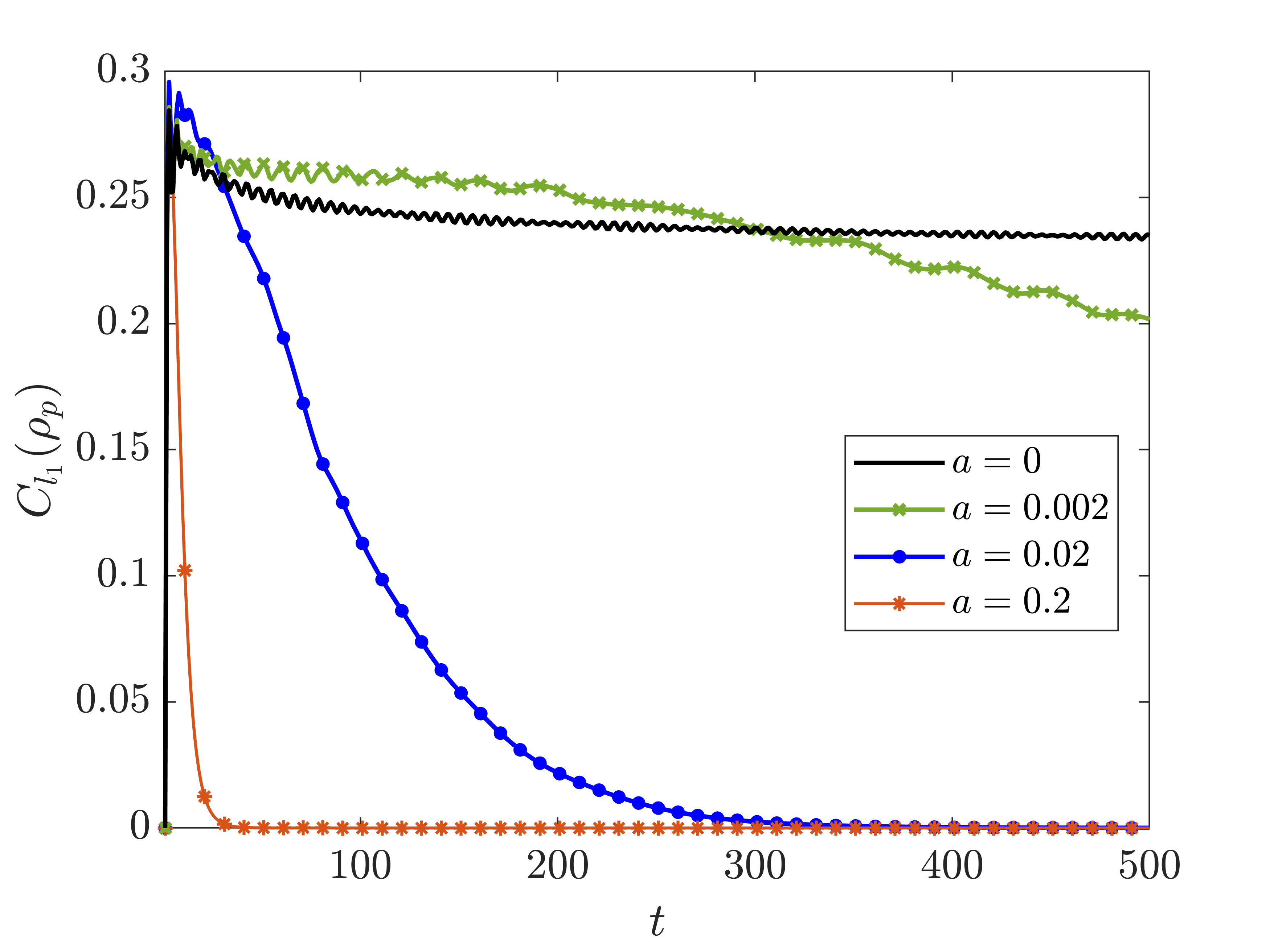}
	\caption{\label{fig:coh_acc_500} $l_1$-norm based coherence from Eq.~\eqref{norm_coh} is determined for walks of increasing acceleration. Evidently, these superdiffusive accelerated walks pose a peculiar argument for coherence compared to homogeneous or disordered walks. Higher the accelerating parameter $a$ in the coin parameter, the faster the coherence measure plummets to zero. The drop is expected to begin once the walk achieves fully ballistic spreading. On the contrary, the coherence in homogeneous walk achieves saturation.}
\end{figure}

As is evident from Fig.~\ref{fig:coherenceL1}, the walks displaying superdiffusive behavior, namely homogeneous and accelerated walks, amass higher coherence values than that of disordered subdiffusive walks. Although all the walks are initialized uniformly, the coherence values show a distinctive behavior in the first few steps themselves. The effect of $1/t$ in Eq.~\eqref{norm_coh} begins to dominate the coherence in walks soon after a few steps.  For quantum walks which returns classical random walk behavior due to subsequent measurement in coin space after every step,  it is straightforward to note that the coherence will be zero.  Due to less spread in position space for disordered quantum walks, coherence is very low compared to the homogeneous and accelerated quantum walks. The decline in the coherence of accelerated walk can be attributed to a decrease in the spread of amplitude in position with an increase in the number of steps. The rate of descent in coherence values is inversely related to the acceleration parameter $a$. It is shown in Fig.~\ref{fig:coh_acc_500} by extending the evolution to 500 steps.  For higher values of $a$, the walker quickly reaches the extreme points $\pm t$ in position space with almost zero probability at all other position space making it a directed transport dynamics, deviating from the anomalous behavior. This decrease in coherence indicates a corresponding decrease in the number of lattice sites on which the walker is in superposition at a given time.

\subsection{Relative entropy of coherence}
The relative entropy of coherence $C_{RE}$ is an induced measure based on the quantum relative entropy $\infdiv{\rho}{\sigma} = \text{Tr}[\rho \log (\rho)] - \text{Tr}[\rho \log (\sigma)]$ for any two states $\rho$ and $\sigma$. This relative entropy measure has been previously used in various contexts such as quantifying superposition~\cite{Abe06, AR15}, estimating frameness~\cite{GMS09, HHH05} and quantum thermodynamics~\cite{VAW08,RFA13}. For a given density matrix $\rho = \sum_{i,j}\rho_{i, j}\ket{i}\bra{j}$, we denote $\rd \equiv \rho_{i,i}\ket{i}\bra{i}$ for a matrix consisting of only its diagonal elements. Reminding that von Neumann entropy is $S(\rho) = -\text{Tr}[\rho \ln(\rho)]$, we define $C_{RE}$ as,
\begin{equation}
\label{coherenceRE}
\mc_{RE}(\rho) = S(\rho_{\text{diag}}) - S(\rho).
\end{equation}
Based on the definition in Eq.~\eqref{coherenceRE} an inequality for obtaining the maximum coherence in a state readily follows: $\mc_{RE} \leq S(\rd) \leq \log d$ with the equality achieved only when the state is maximally coherent. This measure of coherence is also shown to be super-additive and tightly bounded above by the information function for a given quantum state~\cite{XLF15}.

\begin{figure}[!ht]
	\centering
	\includegraphics[width=\columnwidth]{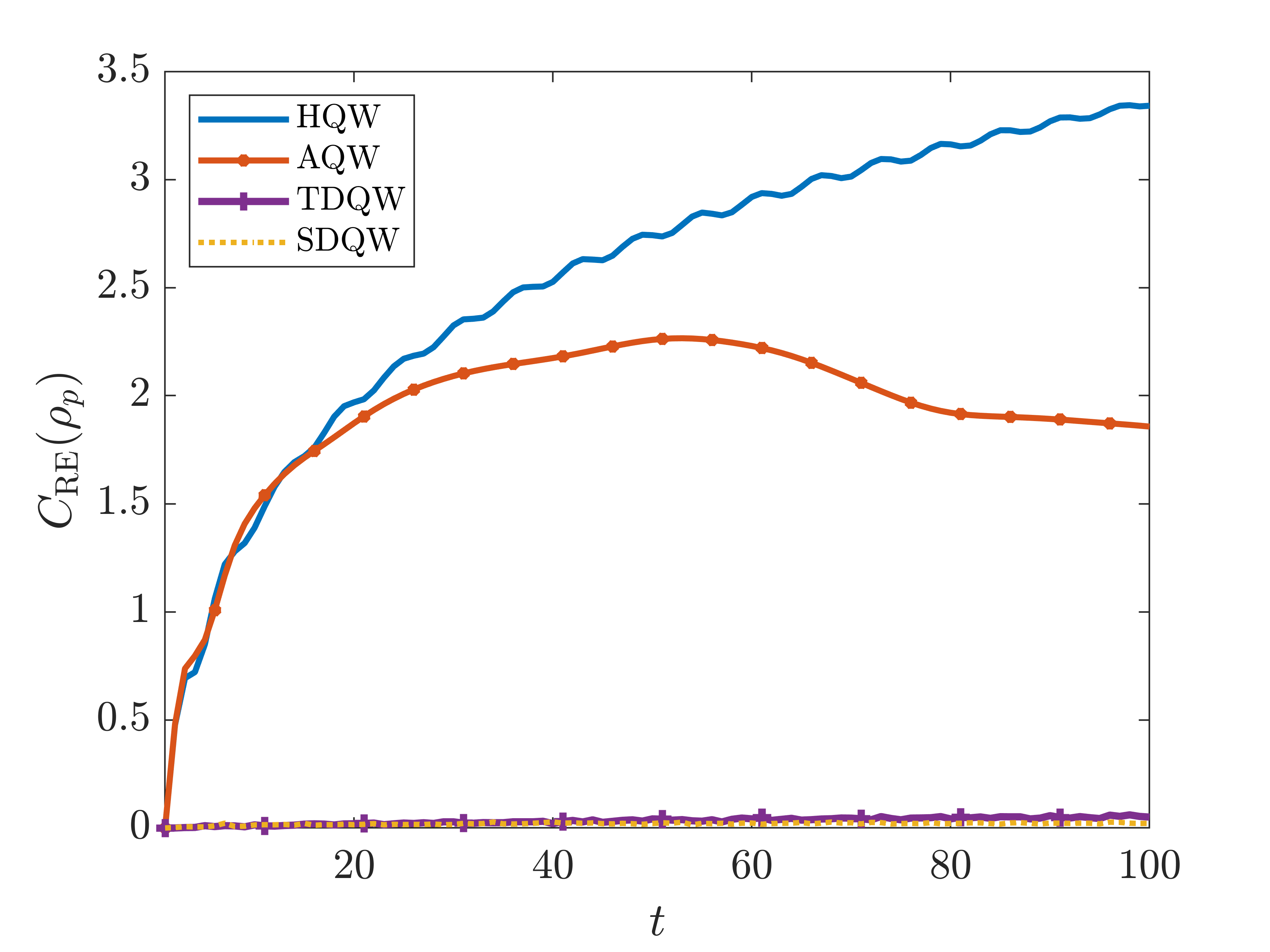}
	\caption{\label{fig:relent_coherence}Relative entropy of coherence as defined in Eq.~\eqref{coherenceRE} is plotted for the positional degree of freedom of different walkers as they evolve in discrete time steps. The specifications used for walks is same as in Fig.~\ref{fig:probability}. The trend in coherence displayed by different walks is strikingly similar to that of Fig.~\ref{fig:coherenceL1}. The superdiffusive walks are a result of highly coherent dynamics while the subdiffusive ones exhibit little to no coherence.}
\end{figure}

\begin{figure}[!ht]
	\centering
	\includegraphics[width=\columnwidth]{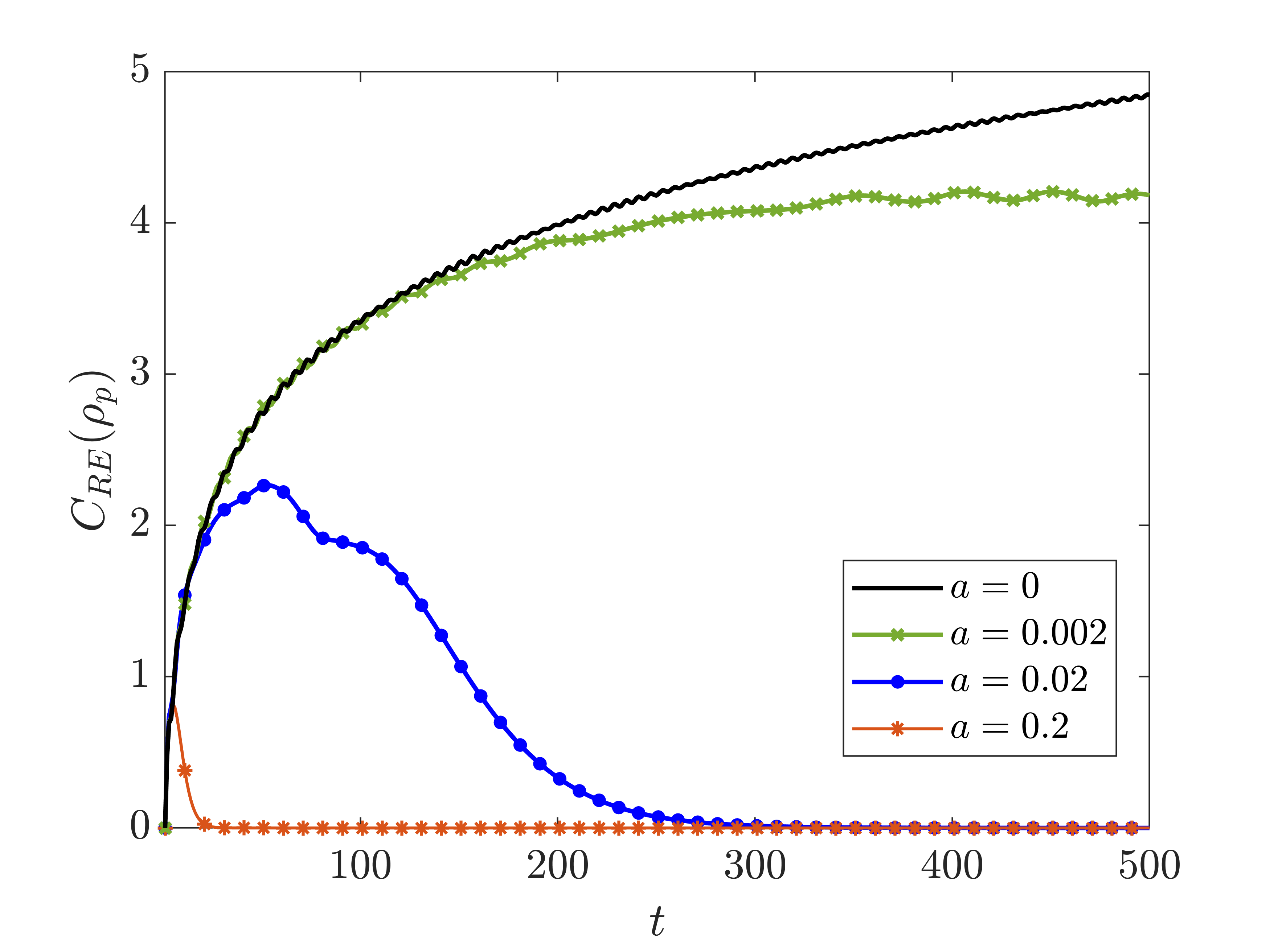}
	\caption{\label{fig:relent_acc_500}Coherence entropy as defined in Eq.~\eqref{coherenceRE} is calculated across the accelerating parameter of increasing order of magnitude. The accelerated walks displaying superdiffusive behavior present a special case of coherence compared to homogeneous or disordered walks. Here as well, the coherence measure is inversely related to the number of steps at which it descends to zero. Even after 500 steps, the homogeneous walk on the other hand steadily increases in coherence.}
\end{figure}

As with the $l_1$-norm based measure of coherence, we focus on the density matrix of external degree of freedom of the walker $\rho_p$ obtained after tracing out the coin Hilbert space. Fig.~\ref{fig:relent_coherence} shows the evolution of relative entropy of the walker with incremental steps. In line with the observations drawn from Fig.~\ref{fig:coherenceL1}, the superdiffusion is correlated with higher coherence values while the disordered walks show non-zero but almost flat coherence curves. Besides pronouncing similar conclusions on interference effects influencing the spread of the walker, the departure of coherence of accelerated walker from that of the homogeneous one proceeds sooner than in Fig.~\ref{fig:coherenceL1}. To probe further into this distinctive behavior of accelerated walks, we evolve the walker for 500 steps in Fig.~\ref{fig:relent_acc_500}. Clearly, the coherence is reaching zero asymptotically for any order of magnitude of the accelerating parameter $a$, whereas the homogeneous walk grows steadily coherent.

\section{Discussion and Summary}
\label{sec:summary}


The main emphasis in the present work has been laid on how interference effectively determines the spreading of a quantum particle modeled using quantum walks. We have demonstrated the versatility of quantum walks resulting in the evolution ranging from superdiffusion in homogeneous and accelerated dynamics to  normal diffusion and subdiffusion in disordered dynamics.  Degree of interference in the dynamics has been quantified using coherence and has been show as an effective way to characterize anomalous diffusion in quantum walks.  From both, $l_1$-norm coherence and coherence entropy  measure we can see that the non-zero value indicate diffusion, the value of $\alpha$ used to characterize the anomalous behavior can be mapped to higher value of coherence when $\alpha > 1$ and lower value when $\alpha < 1$.   Accelerated quantum walk showing the transition from diffusive behavior to transport without interference resulting in decline of coherence value to zero has further established effective use of coherence to  characterize diffusion in quantum walks. The same can be extended to show anomalous diffusion in periodic quantum walks~\cite{PBL18} and quasi-periodic quantum walks~\cite{LAB17} and can be effectively used to characterize anomalous diffusion in quantum dynamics.   Though it is evident from the  study presented here that low value of coherence indicates subdiffusive behavior and high coherence value indicates superdiffusive behavior, the transition from one to the other and characterization of transition using coherence is an interesting question for further probing, particularly from the perspective of open quantum dynamics since it accounts for the evolution of most of the realistic quantum systems. Moreover, the insights furnished by quantum walks will be of immediate relevance in engineering and modeling coherent processes with desired diffusive properties and will be useful in further modeling and characterizing diffusion emerging from the dynamics in various complex quantum systems.

\begin{acknowledgments}
ASH and CMC acknowledge the support from the Office of Principal Scientific Advisor to Government of India, project no. Prn.SA/QSim/2020 and Interdisciplinary Cyber Physical Systems (ICPS) program of the Department of Science and Technology, India, Grant No.: DST/ICPS/QuST/Theme-1/2019/1 for the support.
\end{acknowledgments}


\end{document}